\documentclass[aps,prl,reprint,groupedaddress,showpacs]{revtex4-1}

\usepackage{graphicx}

\newcommand{\EEE}{\mbox{${\cal E}$}}
\newcommand{\LLL}{\mbox{${\cal L}$}}
\newcommand{\SSS}{\mbox{\tiny ${\Sigma}$}}
\newcommand{\Kel}{\mbox{${\cal K}\!e$}}

\begin{document}


\title{Moving contact lines in a pure-vapor atmosphere: \\ a singularity-free description in the sole framework of classical physics}


\author{Alexey Rednikov}
\email[]{aredniko@ulb.ac.be}
\homepage[]{http://www.tips-ulb.be}
\author{Pierre Colinet}
\email[]{pcolinet@ulb.ac.be}
\homepage[]{http://www.tips-ulb.be}
\affiliation{Universit\'e Libre de Bruxelles, TIPs--Fluid Physics, CP 165/67, B-1050 Brussels, Belgium}


\date{\today}

\begin{abstract}
We here show that, even in the absence of ``regularizing'' microscopic effects (viz.\ slip at the wall or the disjoining pressure/precursor films), no singularities in fact arise for a moving contact line surrounded by the pure vapor of the liquid considered. There are no evaporation-related singularities either even should the substrate be superheated. We consider, within the lubrication approximation and a classical one-sided model, a contact line advancing/receding at a constant velocity, or immobile, and starting abruptly at a (formally) bare solid surface with a zero or finite contact angle.  
\end{abstract}


\maketitle

Since the famous paper by Huh \& Scriven \cite{huhscriven}, one is well aware that the {\it moving}-contact-line problem encounters essential difficulties within classical hydrodynamics. The term ``classical'' here includes the no-slip condition at a rigid wall and the treatment of a liquid-gas interface as a geometrical surface, and this is how it is understood hereafter. An attempt at constructing a solution to such a problem \cite{huhscriven} leads to non-integrable, logarithmic divergences of the total drag force and viscous dissipation, as well as the impossibility of satisfying the normal stress condition at the free interface. This forces the incorporation of non-classical, microscopic effects, appreciable just in a tiny vicinity of the contact line, such as the (Derjaguin) disjoining pressure \cite{dGreview,bonneggersrev2009}, resulting in various sorts of precursor films, or a Navier slip at the wall \cite{huhscriven,bonneggersrev2009}. 

\begin{figure} 
\includegraphics[scale=0.55]{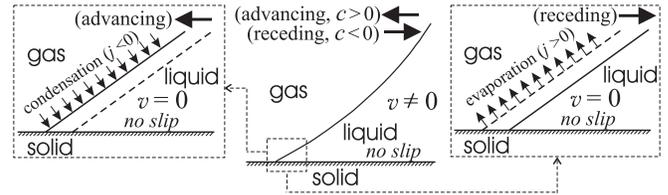}
\caption{Phase change: a possible regularizing effect. While the velocity $v$ in the liquid is generally non-zero ($v\ne 0$), $v\to 0$ towards the contact line, so that the latter advances by condensation and recedes by evaporation (see the zooms).}
\label{idea}
\end{figure}

On the other hand, it seems quite natural to expect that the contact-line motion can somehow be aided by the processes of evaporation and condensation, as rather heuristically studied in \cite{wayner1993}. In \cite{pomeau2011}, it is shown how the Huh \& Scriven singularities can partly be relaxed by the phase change, viz.\ regarding the normal stress balance. Nonetheless, the consideration \cite{pomeau2011} is of a ``kinematic'' character, without specifying how such phase change, exactly of the required rate, could possibly be turned on. Yet, the ``dynamic'' issue is a key here. Indeed, even if one imagines the (kinematic) scenario shown in Fig.~\ref{idea}, it is not a priori clear what would make the phase change adjust itself to the contact line displacement, all the more so that this must be realizable so as to happen for any contact-line velocity (generally determined by the overall macroscopic configuration rather than by what is going on in the vicinity of the contact line) and any contact angle (regarded as a material property of the system -- e.g. the Young's angle), and not just for some specific values. Here we provide a clear-cut model of how this can work, in the case of a contact line in a pure-vapor atmosphere. The goal is to show in principle that the contact-line problem can be resolved classically, and to this end we deliberately keep the number of physical effects to a minimum (``minimalist'' approach). 

\begin{figure} 
\includegraphics[scale=0.7]{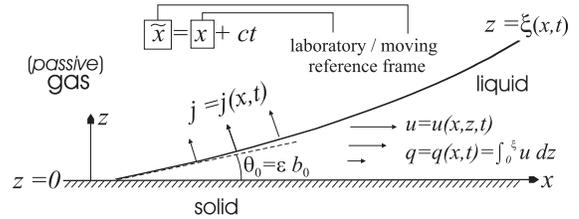}
\caption{Hydrodynamic definition sketch.}
\label{defs}
\end{figure}

On a technical side, we proceed in the classical lubrication (thin-film) approximation, widely employed in the literature for similar problems \cite{dGreview,bonneggersrev2009}, and implying sufficiently small film slopes and contact angles. 
The liquid film is situated on a flat rigid substrate in a planar geometry, see Fig.~\ref{defs}. Gravity and inertia are neglected given the eventual focus upon a small vicinity of the contact line. The horizontal momentum equation reads $0=-\partial p^*/\partial x^*+\mu_l^* \partial^2 u^*/\partial z^{*2}$, where the pressure $p^*=p^*(x^*,t^*)$ is constant along $z^*$ to leading order. Here $t^*$ is the time, and $\mu_l^*$ is the liquid dynamic viscosity. With no slip at the bottom, $u^*=0$ at $z^*=0$, and a stress-free interface, $\partial u^*/\partial z^*=0$ at $z^*=\xi^*$ (neglecting the dynamic influence of the gas), one obtains 
$u^*=\left(1/2 \, z^{*2}-\xi^* z^*\right) \mu_l^{*-1} \partial p^*/\partial x^*$,  
and the volume flux 
\begin{equation}
q^*=\int_0^{\xi^*} u^* dz^*=-\frac{\xi^{*3}}{3\mu_l^*} \frac{\partial p^*}{\partial x^*}
\label{horizflux}
\end{equation}
Let $j^*$ be the local phase-change rate (in ${\rm kg/m^2 s}$, evaporation: $j^*>0$, condensation: $j^*<0$). The volume conservation then reads ($\rho_l^*$ is the liquid density)
\begin{equation}
\frac{\partial\xi^*}{\partial t^*} + \frac{\partial q^*}{\partial x^*}+ \frac{j^*}{\rho_l^*} = 0
\label{filmeq1}
\end{equation}

A closure to the formulation (\ref{filmeq1}) with (\ref{horizflux}) can be in the form of $p^*$ and $j^*$ specified as functionals of $\xi^*$. Here, 
\begin{equation}
p^*=p_0^*-\gamma^* \frac{\partial^2\xi^*}{\partial x^{*2}}
\label{pres}
\end{equation}
as given by the Laplace pressure, where $\gamma^*$ is the surface tension and $p_0^*$ is the gas pressure (whose variation is neglected as compared to that in the liquid, in line with neglecting the dynamic contributions from the gas phase). As for $j^*$, the closure is trivial in the non-volatile case: $j^*=0$. Otherwise, it will be provided later. 

We shall be interested in stationary film profiles (the contact line tip is at $x^*=x_0^*$) translating at a constant velocity $c^*=-dx_0^*/dt^*$ (advancing: $c^*>0$, receding: $c^*<0$). Thus, $\xi^*=\xi^*(\tilde x^*)$ with $\tilde x^*=x^*+c^* t^*$ and we shall take $\tilde x^*=0$ at the contact line itself. Eq.~(\ref{filmeq1}) becomes $c^* d\xi^*/d\tilde x^*+dq^*/d\tilde x^*+j^*/\rho_l^*=0$. Then, 
\begin{equation} 
q^*=-\left(c^* \xi^*+J^*/\rho_l^*\right)
\label{filmeq2}
\end{equation}
with
$J^*=\int_{x_0^*}^{x^*} j^*({x^*} ')\,d{x^*} '$ (assumed to converge at $x_0^*$).

The viscous dissipation and the tangential stress acting on the solid substrate are two quantities often used to put into evidence contact-line singularities: 
\begin{equation}
|\dot{\EEE}_A^*|=\mu_l^*\int_0^{\xi^*} \left(\frac{\partial u^*}{\partial z^*}\right)^2 dz^*=3\mu_l^*\frac{q^{*2}}{\xi^{*3}}
\label{dissip}
\end{equation}
\vspace{-0.2in}
\begin{equation}
\sigma_{\tau}^*=\mu_l^* \frac{\partial u^*}{\partial z^*}\Bigg|_{z^*=0}=3\mu_l^*\frac{q^*}{\xi^{*2}}
\label{tangstr}
\end{equation}
where the dissipation density $|\dot{\EEE}_A^*|$ is already expressed per unit area of the film. 

Consider first the non-volatile case ($j^*=0$, $J^*=0$) and revisit the classical singularities \cite{dGreview} in the present formulation. For a finite contact angle, $\theta_0= d\xi^*/d\tilde x^*|_{\tilde x^*\to 0}>0$, we have $\xi^*\sim \tilde x^*$. Thus, from Eq.~(\ref{filmeq2}) (with $c^*\neq 0$, $J^*=0$), $q^*\sim \tilde x^*$. Now from Eqs.~(\ref{dissip}) and (\ref{tangstr}), $|\dot{\EEE}_A^*|\sim \tilde x^{*-1}$ and $\sigma_{\tau}^*\sim \tilde x^{*-1}$, diverging non-integrably as $\tilde x^*\to 0$. 
Moreover, from (\ref{horizflux}), one establishes $p^*\sim \tilde x^{*-1}$. But then from (\ref{pres}), one obtains the correction $\xi^*\sim\tilde x^* \log \tilde x^*$ over the supposed leading-order film-thickness behavior $\xi^*\sim\tilde x^*$ (a finite contact angle) as $\tilde x^*\to 0$, the former contradictorily exceeding the latter. 
Note that singularities become even more unbearable should one attempt to consider a zero contact angle. 

On its own, the phase change is potentially as much a source of singularities as the motion is, and perhaps even more so, even though this seems to be less widely recognized. For instance, hydrodynamic singularities even stronger than those considered above are expected in the case of the phase-change-flux behavior $j^*\sim \tilde x^{*-1/2}$ taking place at the edge of a thin sessile drop evaporating into an inert gas \cite{deegan_nature1997} (they would be the same as above for $j^*\sim \tilde x^{*0}$). In the present Letter, the issue of such evaporation-induced singularities is addressed as well. 

However it may be, from Eqs.~(\ref{filmeq2})--(\ref{tangstr}), we see that the singularities {\it can} be mitigated should both the factors (the motion and phase change) act together in the right way: $c^*\xi^*+J^*/\rho_l^*=o(\tilde x^*)$ as $\tilde x^*\to 0$. Equivalently, 
\begin{equation}
j^*\to -\rho_l^* c^* \frac{d\xi^*}{d\tilde x^*} = -\rho_l^* c^* \theta_0 \quad {\rm as}\  \tilde x^*\to 0
\label{necescond}
\end{equation}
whose interpretation is straightforward, and actually corresponds to the scenario already imagined in Fig.~\ref{idea}. 

Nonetheless, as emphasized before, it is not a priori guaranteed that the condition (\ref{necescond}) can actually be realized in a concrete system or, what matters for the purposes of the present Letter, that one can provide a theoretical example thereof. The point is that we cannot just specify $j^*$ at will so as to satisfy (\ref{necescond}). The system must have mechanisms to fine-tune itself in such a way on its own. This must also be flexible, i.e. for a range of values of $c^*$ and $\theta_0$, and not just for some degenerate ones. 

Now let the gas phase be the pure vapor of the liquid. In the simplest case, the temperature $T_w^*$ of the wall is just equal to the saturation temperature $T_0^*\equiv T_{\rm sat}^*(p_0^*)$ for the vapor pressure $p_0^*$. This case is already quite sufficient for our goals as far as motion-induced singularities are concerned (cf.\ below). If in spite of our minimalist strategy we nonetheless consider a more general case, with a nonzero superheat $\Delta T^*\equiv T_w^*-T_0^*$ (as e.g.\ in boiling applications), this is because it is equally of interest to illustrate the genericity of our approach also as far as evaporation-induced singularities are concerned (be it for immobile contact lines), which, as already mentioned, are a big issue as well. The phase change flux is principally determined by heat conduction through the liquid film \cite{BBD1988}, and consistent with the thin-film approximation used throughout this Letter one simply obtains
\begin{equation}
j^*=\LLL^{*-1}\lambda_l^* (T_w^*-T_{\SSS}^*)/\xi^*
\label{evapflux}
\end{equation}
where $T_{\SSS}^*$ is the liquid-vapor interface temperature, $\lambda_l^*$ is the thermal conductivity of the liquid, and $\LLL^*$ is the latent heat of evaporation (energy per unit mass).
With local equilibrium at the liquid-vapor interface, we have $T_{\SSS}^*=T_0^*$. For a non-zero superheat, the flux (\ref{evapflux}) then turns out to be non-integrably singular as $\tilde x^*\to 0$ where $\xi^*\sim \tilde x^*$. This thermal singularity can be relaxed, on the one hand, by the effect of finite-rate phase-change kinetics (typically very small and limited to a microscopic vicinity of the contact line), and on the other hand, 
by a finite (rather than formally infinite as assumed here) thermal conductivity of the wall \cite{sadhalplesset1979}, when $T_w^*\ne {\rm const}$ in space and $T_w^*=T_0^*$ at the contact line itself. However, these two effects can hardly possibly account for the earlier mentioned fine-tuning (\ref{necescond}), all the more so that they are evidently useless in this regard for zero superheat. Staying minimalist, we shall not take them into account. Another effect typically recognized as important within the microstructure of liquid-vapor contact lines \cite{wayner1993} (see also \cite{papier0} and references therein) is the Kelvin effect, according to which, in our context, the saturation temperature is no longer $T_0^*$ but rather
$T_{0,{\rm loc}}^*=T_0^*+\left(T_0^*\gamma^*/\LLL^*\rho_l^*\right) \partial^2\xi^*/\partial x^{*2}$, 
varying along the film together with the curvature $\partial^2\xi^*/\partial x^{*2}$. Here the deviations from $T_0^*$ are assumed small. Now $T_{\SSS}^*=T_{0,{\rm loc}}^*$ in (\ref{evapflux}). The relaxation of the thermal singularity may then be possible if the film curvature $\partial^2\xi^*/\partial x^{*2}$ self-adjusts itself towards the contact line at a value $(\LLL^*\rho_l^*/\gamma^*)(\Delta T^*/T_0^*)$. We shall see below that this Kelvin-effect mechanism automatically takes care of the hydrodynamic singularities as well, which will also work for zero superheat. Its being determined by the second derivative $\partial^2\xi^*/\partial x^{*2}$ (coming at a higher order than the contact-line slope itself as $\tilde x^*\to 0$) is eventually what makes such a subtle regulation possible. Note though that very small radii of curvature are typically required for the Kelvin effect to be essential. This is what actually gives rise to the contact line possessing a microstructure (a microregion with the corresponding microscales) even within a classically-based treatment as ours. On the other hand, on the macroscopic scale, the Kelvin effect becomes negligible and the simplest formulation is recovered, as expected. 

With $T_{\SSS}^*=T_{0,{\rm loc}}^*$ in (\ref{evapflux}), and substituting (\ref{horizflux}), (\ref{pres}) and (\ref{evapflux}) into (\ref{filmeq1}), we obtain the final film equation. We still consider the case $\xi^*=\xi^*(\tilde x^*)$ with $\tilde x^*=x^*+c^* t^*$, and are interested in solutions emanating from a bare solid surface at a given (Young's) contact angle $\theta_0$ ($\xi^*\sim \theta_0 \tilde x^*$ as $\tilde x^*\to 0$) towards $\tilde x^*>0$. At this stage, it is convenient to render the formulation dimensionless.  
An important convention: for any quantity $f$ we write $f^*=[f]\,f$, where the asterisk or its absence distinguishes between dimensional and dimensionless versions, respectively, and $[f]$ is the scale. We then obtain   
\begin{equation}
c \,\xi'+\left(\xi^3\xi'''\right)'+\left(E-3\Kel\,\xi''\right)/\xi=0
\label{filmeq3}
\end{equation}
\vspace{-0.35in}
\begin{equation}
\xi\sim b_0 \,\tilde x\quad {\rm as}\ \tilde x\to 0
\label{anglebc}
\end{equation}
\vspace{-0.3in}
$$c=\frac{3 C\!a}{\epsilon^3}\,,\ E=\frac{3 C\!a_{\rm evap}}{\epsilon^4}\,,\ \Kel=\frac{\mu_l^*\lambda_l^* T_0^*}{\rho_l^{*2}\LLL^{*2}[\xi]^2 \epsilon^2}\,,\ b_0=\frac{\theta_0}{\epsilon}
$$
\vspace{-0.25in}
$$\epsilon=\frac{[\xi]}{[x]}\,,\ C\!a=\frac{\mu_l^* c^*}{\gamma^*}\,,\ 
C\!a_{\rm evap}=\frac{\mu_l^*\lambda_l^*\Delta T^*}{\gamma^*\rho_l^*\LLL^* [\xi]}
$$
where the prime denotes a derivative with respect to $\tilde x$ ($[\tilde x]=[x]$), and  $[c]=[u]=\gamma^*\epsilon^3/3\mu_l^*$. The use of numerical coefficients in definitions of  dimensionless numbers is partly due to traditions (cf.\ e.g.\ \cite{papier0}). Generally, $c=O(1)$, $E=O(1)$, $\Kel=O(1)$, and $b_0=O(1)$. However, for the thin-film approximation to be valid, one needs $\epsilon\ll 1$, and hence $C\!a\ll 1$, $C\!a_{\rm evap}\ll 1$, $\theta_0\ll 1$. We have not yet specified $[\xi]$ and $[x]=[\xi]/\epsilon$. In doing so, the number of parameters in (\ref{filmeq3})-(\ref{anglebc}) is reduced by two. As the Kelvin effect is the key here, $\Kel\equiv 1$ would be a reasonable start, whereas the second relation could be $|c|\equiv 1$ (if the contact line motion is important), or $E\equiv 1$ (if the superheat is important), or rather $b_0\equiv 1$ (if the motion and evaporation do not modify the film slopes significantly). 

Using Eq.~(\ref{filmeq3}), 
the coordinate expansion behind (\ref{anglebc}) is 
$$\hspace{-0.9in}\xi=b_0 \tilde x+ \frac{E}{6\Kel} \,\tilde x^2 + \frac{b_0^2 c}{18 \Kel} \,\tilde x^3 + \frac{b_0 c E}{72 \Kel^2} \,\tilde x^4$$
\vspace{-0.3in}
\begin{equation}
\hspace{0.0in}\mbox{}+
\frac{\left(E^2+4 b_0^3 c\Kel+18 b_0^6\Kel\right) c}{1080 \Kel^3} \,\tilde x^5 +O(\tilde x^6)+\xi_{\rm eigen}\ {\rm as}\ \tilde x\to 0
\label{asyini}
\end{equation}
where $\xi_{\rm eigen}$ is an exponentially decaying eigenfunction contribution discussed below. Eq.~(\ref{asyini}) is 
valid in the case of a finite contact angle ($b_0>0$, partial wetting) unconditionally, i.e.\ for advancing ($c>0$), receding ($c<0$) or immobile ($c=0$) contact lines, and for positive ($E>0$), zero ($E=0$) or even negative ($E<0$) superheats. For a zero contact angle ($b_0=0$, perfect wetting), however, the validity of (\ref{asyini}) is limited to positive superheats ($E>0$), but still without limitations on $c$. Indeed, for $b_0=0$ and $E=0$, Eq.~(\ref{asyini}) degenerates. As for $b_0=0$ and $E<0$, Eq.~(\ref{asyini}) yields a clearly unphysical result (negative film thicknesses), which is deemed to be an indication that the film structures studied here (with a contact line) simply do not exist in this case. Anyhow, it can readily be appreciated that no contact-line singularities are actually associated with the solution (\ref{asyini}), e.g.\ by using the dimensional version of it in (\ref{pres}), and then in (\ref{horizflux}), (\ref{dissip}), (\ref{tangstr}). Furthermore, the scenario anticipated in Fig.~\ref{idea} and Eq.~(\ref{necescond}) can be seen to be fully realized. This is the principal result of this Letter. Eq.~(\ref{asyini}) confirms the key role of the Kelvin effect, for there exists no limit $\Kel\to 0$. 

Now $\xi_{\rm eigen}$, which can be obtained by linearizing Eq.~(\ref{filmeq3}) near the algebraic part of (\ref{asyini}), reads  
$$\xi_{\rm eigen}=B\,\exp\left(-\frac{\sqrt{3 \Kel}}{b_0^2 \tilde x}\right)\,\tilde x^{\beta} \bigg(1+ O(\tilde x)\bigg)\quad {\rm as}\ \tilde x\to 0
$$
with $\beta=7/2-E\,b_0^{-3}/\sqrt{3\Kel}$ for $b_0>0$, and 
$$\xi_{\rm eigen}=B\, \exp\left(-\frac{12\sqrt{3\Kel^5}}{E^2 \tilde x^3}\right)\,\tilde x^{\beta} \bigg(1+O(\tilde x^3)\bigg)\quad {\rm as}\ \tilde x\to 0
$$
with $\beta=7-2\sqrt{3\Kel}\, c/5E$ for $b_0=0$ (only $E>0$). The arbitrary coefficient $B$ makes it plausible that a solution to the problem (\ref{filmeq3}) with (\ref{anglebc}) can exist for quite a class of possible boundary conditions to the right (posed at some $\tilde x>0$). For instance, one can be interested in solutions with a given non-negative value of the curvature $\xi''$ far away from the present Kelvin-effect-induced microstructure 
(formally, as $\tilde x\to +\infty$). For such a boundary-value problem, $B$ will be an eigenvalue. 

The already mentioned degeneracy of (\ref{asyini}) for $b_0=0$, $E=0$ signals that no actual contact line may exist (on the microscopic scale) in the perfect-wetting case and without superheat. This does not seem to be counterintuitive, all the more so that a ``topologically close'' (cf.\ also later) structure can be found instead: 
$$\hspace{-0.5in} \xi=-\frac{6\,\Kel}{c\,\tilde x} + O(\tilde x^{-7}) + B \exp\left(\frac{c^2 \tilde x^3}{36\sqrt{3\Kel^3}}\right)$$
\vspace{-0.3in}
\begin{equation}
\hspace{0.5in}\mbox{}\times (-\tilde x)^{-5/2}\left(1+O(\tilde x^{-3})\right)     \quad {\rm as}\ \tilde x\to -\infty
\label{asymaxi}
\end{equation}
valid for an advancing contact line ($c>0$) without superheat ($E=0$). Eq.~(\ref{asymaxi}) describes a microfilm (precursor film) extending over the solid surface ahead of the macroscopic portion of the liquid, with a thickness asymptotically decaying far away. De~Gennes and collaborators \cite{dGreview} referred to such type of structure as a maximal solution, even though note that its physical origin is quite different here and there. There, it was due to the disjoining pressure and the liquid was non-volatile. Here, it is due to the Kelvin effect and the liquid is volatile. On the other hand, a receding ($c<0$) contact line in the perfect-wetting case ($b_0=0$) without superheat ($E=0$) must leave behind itself a constant-thickness film, much like in the Landau-Levich problem (see e.g.\ \cite{bonneggersrev2009}), and this case will be studied in more detail elsewhere. 

Above, we have focused on the asymptotic behavior at the start of the film, for after all the main goal of this Letter is to make sure that everything is alright there from the viewpoint of singularities. For an illustration of the contact-line microstructure as a whole, we shall consider a ``quasi-wedge'' problem such that 
\begin{equation}
\xi\sim \tilde x\,\left(3 c \log \tilde x + b_{\rm eff}^3\right)^{1/3}\quad {\rm as}\ \tilde x\to +\infty
\label{asyinf}
\end{equation}
which is compatible with Eq.~(\ref{filmeq3}), valid for $c\ge 0$ (i.e.\ either an advancing or an immobile contact line) and represents a Tanner-Cox-Voinov-like behavior \cite{bonneggersrev2009}. Eq.~(\ref{asyinf}) obviously implies that $\xi''\to 0$ as $\tilde x\to +\infty$, which is an idealization of the fact that the macroscopic curvature can be much smaller than the microscopic one in the Kelvin-effect-induced microregion. The boundary-value problem is then given by Eq.~(\ref{filmeq3}) with (\ref{asyini}) or (\ref{asymaxi}) and (\ref{asyinf}), in which $b_{\rm eff}$ is an eigenvalue. We have intentionally chosen to express the free coefficient in the asymptotics (\ref{asyinf}) in such a form (with $b_{\rm eff}$) looking forward to the fact that, for $c=0$, $b_{\rm eff}$ will be just the (rescaled with $\epsilon$) apparent contact angle (different from $b_0$) as seen on the macroscale. For $c\neq 0$, such an apparent contact angle will rather be $b_{\rm app}=(3c\log l+ b_{\rm eff}^3)^{1/3}$, or in the original (non-rescaled) terms $\theta_{\rm app}=(9 C\!a\log l+ \theta_{\rm eff}^3)^{1/3}$, where $l\gg 1$ is a dimensionless macroscopic scale and $\theta_{\rm eff}= \epsilon b_{\rm eff}$, the result being applicable in such a context (with a large but finite $l$, unlike (\ref{asyinf})) even for receding contact lines ($c<0$) provided that $b_{\rm eff}$ is sufficiently larger than $(3 |c| \log l)^{1/3}$. Coming back to our boundary-value problem, we shall limit ourselves to two distinguished cases (`a' and `b'), the (numerical) results for which are shown in Fig.~3 (while a more complete parametric study will be published elsewhere). The scaling is here concretized as discussed below Eqs.~(\ref{filmeq3}) and (\ref{anglebc}). Note that for both `a' and `b', $b_{\rm eff}$ approaches $b_0$ for large values of the latter, in which limit the motion- and evaporation-induced slopes (respectively) are small relative to the Young's angle. For `a', it is seen that the maximal solution is the limit attained by the film profiles as $b_0$ is decreased. For a finite superheat (case `b'), however, this limit ($b_0\to 0$) corresponds to a profile with a finite starting point, as foreseen above from coordinate expansions. For `b', the evaporation-induced apparent contact angle is non-zero even for $b_0=0$, as expected (cf.\ e.g.\ \cite{papier0}). We conclude by noting that the present singularity-free approach can be extended to include other effects (a bounded disjoining pressure isotherm, finite-rate kinetics, Navier slip, ...) without major difficulties. 

\begin{figure} 
\includegraphics[scale=0.7]{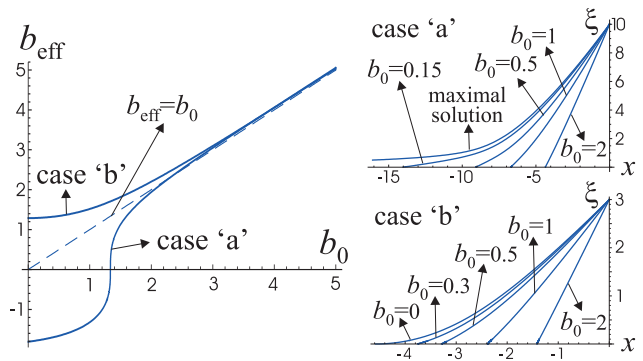}
\caption{$b_{\rm eff}$ versus $b_0$ and the typical starting film profiles (shifted along $x$ so as to pass through the same point to the right) for a) an advancing contact line in the absence of superheat ($E=0$, $c\equiv 1$, $\Kel\equiv 1$), and b) an immobile contact line on a superheated substrate ($c=0$, $E\equiv 1$, $\Kel\equiv 1$).}
\label{paramstudy}
\end{figure}

Financial support of ESA/BELSPO-PRODEX, EU-MULTIFLOW \& FRS-FNRS is gratefully acknowledged. 

\bibliography{no-sing_RednCol_biblio}

\end{document}